\documentclass[12pt]{iopart}

\usepackage{graphicx}

\newcommand{\pra}{Physical Review A}
\newcommand{\aap}{A\&A}
\newcommand{\apj}{ApJ}

\usepackage{iopams}  
\begin{document}

\title[]{Numerical solution of the linear dispersion relation in a
  relativistic pair plasma}

\author{J P\'etri and J G Kirk}

\address{Max-Planck-Institut f\"ur Kernphysik, Saupfercheckweg 1,
  69117 Heidelberg, Germany} \ead{j.petri@mpi-hd.mpg.de}

\begin{abstract}
%  Relativistic plasmas, in which the streaming and/or thermal speeds
%  of the constituent particles are comparable to the speed of
%  light~$c$ play an important role in astrophysical objects such as
%  gamma-ray bursts, jets and pulsar winds.  In order to understand the
%  nature of these objects it is, therefore, necessary to study the
%  various instabilities which are present in, for example, current
%  sheets and shock fronts in such plasmas.  Because relatively few
%  analytic results are available, 
  We describe an algorithm that computes the linear dispersion relation
  of waves and instabilities in relativistic plasmas within a
  Vlasov-Maxwell description.  The method used is fully relativistic
  and involves explicit integration of particle orbits along the
  unperturbed equilibrium trajectories. We check the algorithm against
  the dispersion curves for a single component magnetised plasma and
  for an unmagnetised plasma with counter-streaming components in the
  non-relativistic case. New results on the growth rate of the Weibel
  or two-stream instability in a hot unmagnetised pair plasma
  consisting of two counter-streaming relativistic Maxwellians are
  presented. These are relevant to the physics of the relativistic
  plasmas found in gamma-ray bursts, relativistic jets and pulsar
  winds.
\end{abstract}

\submitto{\PPCF}

\maketitle

\section{Introduction}

Relativistic shock fronts and currents sheets in relativistic flows
play an important role in astrophysical models of gamma-ray bursts
(for a review see Piran~\cite{2005RvMP...76.1143P}), of jets and of
pulsar winds (see for instance Michel~\cite{2005RMxAC..23...27M} and
Kirk~\cite{2005PPCF...47B.719K}). The underlying plasma is probably
composed of electrons, positrons and protons, whose temperature may be
relativistic, i.e. comparable to their rest mass energy.
  
We present an algorithm that computes the linear dispersion relation
of waves in such plasmas within a Vlasov-Maxwell description. The
method used is based on that presented by
Daughton~\cite{1999PhPl....6.1329D} for non relativistic Maxwellians,
and involves explicit time integration of particle orbits along the
unperturbed trajectories. We modify and extend this method by changing
the manner in which the roots of the dispersion relation are located
and adopt a fully relativistic approach, i.e. relativistic
temperatures as well as relativistic drift speeds.

Particular emphasis is given to the two-stream or Weibel
instability~\cite{1959PhRvL...2...83W}. This instability is very
important in astrophysical processes because it is able to generate a
magnetic field by pumping free energy from the anisotropic momentum
distribution of an unmagnetised plasma or from the kinetic drift
energy. There is an extensive literature about the Weibel instability.
Although the dispersion curves has been found in some special cases
such as, for example a water-bag distribution function,
(Yoon~\cite{1987PhRvA..35.2718Y}), or a fully relativistic
bi-Maxwellian distribution function,
(Yoon~\cite{1989PhFlB...1.1336Y}), looking for an analytical
expression for the dispersion relation for a given equilibrium
distribution function is a complicated or even impossible task.  The
water-bag distribution is also the preferred profile to analyse
magnetic field generation in fast ignitor scenarios, (Silva et
al.~\cite{2002PhPl....9.2458S}) or in relativistic shocks, (Wiersma
and Achterberg~\cite{2004A&A...428..365W}, Lyubarsky and
Eichler~\cite{2006ApJ...647.1250L}). The Weibel instability in a
magnetised electron-positron pair plasma has been investigated by Yang
et al~\cite{1993PhFlB...5.3369Y} for the water-bag and for a smooth
distribution function and a covariant description has been formulated
by Melrose~\cite{1982AuJPh..35...41M} and by
Schlickeiser~\cite{2004PhPl...11g5532S}. General conditions for the
existence of the relativistic Weibel instability for arbitrary
distribution functions are discussed in~\cite{2006PhPl...13b2107S}.
Wave propagation in counter-streaming magnetised nonrelativistic
Maxwellian plasmas are studied
in~\cite{2005PhPl...12.2901T,2006PhPl...13f2901T}.  The stability
properties of a nonrelativistic Harris current sheet have been studied
by Daughton~\cite{1999PhPl....6.1329D} and also by Silin et
al.~\cite{2002PhPl....9.1104S}. Streaming instabilities in
relativistic magnetised plasmas and superluminous wave propagation are
discussed by Buti~\cite{1970PhRvA...1.1772B,1972PhRvA...5.1558B}.

In this paper, we investigate the relativistic Weibel instability for
two counter-streaming Maxwellian distribution functions in an
unmagnetised pair plasma. Our approach follows that of Zelenyi et
al.~\cite{1979SvA....23..460Z}, who studied the tearing mode
instabilities in a relativistic Harris current sheet, by integrating
first order perturbations of the relativistic Maxwellian distribution
function along unperturbed particle trajectories. Our algorithm is
compared and checked against the dispersion curves for a single
component magnetised plasma and for an unmagnetised plasma with
counter-streaming components in the non-relativistic case. We present
results for the growth rate of the Weibel instability in a hot
unmagnetised pair plasma consisting of counter-streaming relativistic
Maxwellians.

The paper is organised as follows. In Sec.~\ref{sec:Equilibre}, we
present the full set of non-linear equations governing the motion in
the pair plasma and describe the equilibrium of the two
counter-streaming relativistic Maxwellians. Next, in
Sec.~\ref{sec:linearisation}, the full set of linearised
Vlasov-Maxwell equations for the perturbed electromagnetic potentials
around the equilibrium state are presented.  The eigenvalue problem
and the algorithm are discussed in Sec.~\ref{sec:eigenvalue}. Results
for magnetised plasma wave oscillations and for the non-relativistic
Weibel instability as well as for the relativistic Weibel instability
are shown in Sec.~\ref{sec:results}.  Conclusions are drawn in
Sec.~\ref{sec:conclusion}.

\section{Equilibrium}
\label{sec:Equilibre}

Our purpose is to study the Weibel instability in a relativistic
plasma. This plasma is made of counter-streaming electrons and
positrons with relativistic temperatures and evolving in a static
external magnetic field aligned with the z-axis such that
\begin{equation}
  \label{eq:ChampMag}
  \vec{B}_0 = B_0(x) \, \vec{e}_\mathrm{z}
\end{equation}
In equilibrium, there is no electric field, $\vec{E}_0 = \vec{0}$ and
the charges drift in the y~direction at a relativistic velocity~$\pm
U_s$, ($+$~for positrons and $-$~for electrons with~$U_s>0$). We use
Cartesian coordinates, denoted by~$({\rm x,y,z})$, and the
corresponding basis~$(\vec{e}_\mathrm{x}, \vec{e}_\mathrm{y},
\vec{e}_\mathrm{z})$.  The distribution function at equilibrium for
each species~''$s$'', denoted by $f_{0s}(\vec{r},\vec{p})$, is assumed
to be a relativistic Maxwellian with drift speed~$\pm U_s$. Adopting
the usual notations, namely $t, \vec{r}, \vec{v}, \vec{p}, m_s, q_s$
for respectively the time, position, 3-velocity, 3-momentum, mass and
charge of a particle of species~$s$, the stationary distribution
function reads~:
\begin{equation}
  \label{eq:FDD}
  f_{0s}(\vec{r},\vec{p}) = \frac{n_{0s}(\vec{r})}
  {4 \, \pi \, m_s^3 \, c^3 \, \Theta_s \, K_2(1/\Theta_s)}
  \, \mathrm{exp}[-\Gamma_s \, ( E - U_s \, p_\mathrm{y} ) / \Theta_s \, m_s \, c^2]
\end{equation}
$n_{0s}(\vec{r})$ is the particle number density, $E$ the total energy
of a particle, $p_\mathrm{y}$ the y-component of its momentum, $c$ the
speed of light, $\Gamma_s = 1 / \sqrt{ 1 - U_s^2 / c^2}$ the Lorentz
factor associated with the drift motion and $K_2$ the modified Bessel
function of the second kind. The temperature of the gas is normalised
to the rest mass energy of the leptons such that
\begin{equation}
  \label{eq:Theta}
  \Theta_s = \frac{k_B \, T_s}{m_s \, c^2}
\end{equation}
and $k_B$ is the Boltzmann constant. We introduce the standard
electromagnetic scalar and vector potentials~$(\phi,\vec{A})$, related
to the electromagnetic field~$(\vec{E}, \vec{B})$ by~:
\begin{eqnarray}
  \label{eq:Potentiel}
  \vec{E} & = & - \vec{\nabla} \phi - \frac{\partial\vec{A}}{\partial t} \\
  \vec{B} & = & \vec{\nabla} \wedge \vec{A}
\end{eqnarray}
We employ the Lorenz gauge condition by imposing
\begin{equation}
  \label{eq:Jauge}
  \mathrm{div}\, \vec{A} + \varepsilon_0 \, \mu_0 \, \frac{\partial\phi}{\partial t} = 0
\end{equation}
with $\varepsilon_0 \, \mu_0 \, c^2 = 1$. The relation between
potentials and sources then reads~:
\numparts
\begin{eqnarray}
  \label{eq:OndePhi}
  \Delta \phi - \frac{1}{c^2} \, \frac{\partial^2\phi}{\partial t^2} +
  \frac{\rho}{\varepsilon_0} & = & 0 \\
  \label{eq:OndeA}
  \Delta \vec{A} - \frac{1}{c^2} \, \frac{\partial^2\vec{A}}{\partial t^2} + 
  \mu_0 \, \vec{j} & = & 0
\end{eqnarray}
\endnumparts
The source terms represented by the charge~$\rho$ and
current~$\vec{j}$ densities, are expressed in terms of the
distribution functions by~:
\numparts
\begin{eqnarray}
  \label{eq:SourceRho}
  \rho(\vec{r},t) & = & \sum_s q_s \, \int\!\!\!\int\!\!\!\int 
  f_s(\vec{r},\vec{p},t) \, \rmd^3\vec{p} \\
  \label{eq:SourceJ}
  \vec{j}(\vec{r},t) & = & \sum_s q_s \, \int\!\!\!\int\!\!\!\int 
  \frac{\vec{p}}{\gamma\,m_s} \, f_s(\vec{r},\vec{p},t) \, \rmd^3\vec{p}
\end{eqnarray}
\endnumparts
where $\gamma = \sqrt{1+p^2/m_s^2\,c^2}$ is the Lorentz factor of a
particle.  The time evolution of the distribution functions~$f_s$ is
governed by the well-known relativistic Vlasov-Maxwell equations
written for each species~:
\begin{equation}
  \label{eq:Vlasov}
  \frac{\partial f_s}{\partial t} + \vec{v} \cdot \frac{\partial f_s}{\partial \vec{r}}
  + q_s \, ( \vec{E} + \vec{v} \wedge \vec{B} ) \,
  \cdot \frac{\partial f_s}{\partial \vec{p}} = 0
\end{equation}
The self-consistent non-linear evolution of the plasma is entirely
determined by the set of
equations~(\ref{eq:Potentiel})-(\ref{eq:Vlasov}).

\section{Linearisation}
\label{sec:linearisation}

Our next step is to investigate the stability properties of the
equilibrium configuration given in~(\ref{eq:FDD}). The Vlasov
equation~(\ref{eq:Vlasov}) is linearised for each species about the
equilibrium state~$f_{0s}$, (\ref{eq:FDD}), to obtain the time
evolution of the first order perturbation as
\begin{equation}
  \label{eq:Lineaire}
  \frac{\rmd f_{1s}}{\rmd t} = - q \, ( \vec{E}_1 + \vec{v} \wedge \vec{B}_1 ) 
  \, \cdot \frac{\partial f_{0s}}{\partial \vec{p}}
\end{equation}
Perturbations are denoted by the subscript~$1$ whereas the equilibrium
quantities are depicted by a subscript~$0$. The total time
derivative~$\rmd/\rmd t$ is to be taken along the trajectories of the
particles in the unperturbed electromagnetic field~$(\vec{E}_0 =
\vec{0}, \vec{B}_0)$. More explicitly, performing the time integration
in~(\ref{eq:Lineaire}), the perturbed distribution function is given
by
\begin{equation}
  \label{eq:f1s}
  f_{1s}(\vec{r},\vec{p},t) = - q \, \int_{-\infty}^t 
  \left[ \vec{E}_1(\vec{r}\,',t') + \vec{v}\,' \wedge \vec{B}_1(\vec{r}\,',t') \right] 
  \, \cdot \frac{\partial f_{0s}}{\partial \vec{p}}(\vec{r}\,',\vec{p}\,') \, \rmd t'
\end{equation}
We assume that the perturbation of the distribution function at
initial time vanishes, i.e., $f_{1s}(\vec{r}, \vec{p}, t=-\infty) =
0$. Position and momentum are determined by solving the equations of
motion for a single particle in the unperturbed electromagnetic
field~$(\vec{E}_0 = \vec{0}, \vec{B}_0)$. Thus, the unperturbed
trajectories are solutions of the following system of ordinary
differential equations
\numparts
\begin{eqnarray}
  \label{eq:intmvtobsr}
  \frac{\rm d\vec{r}\,'}{\rmd t'} & = & \vec{v}\,'(t') \\
  \label{eq:intmvtobsp}
  \frac{\rm d\vec{p}\,'}{\rmd t'} & = & q \, \vec{v}\,'(t') \wedge \vec{B}_0(\vec{r}\,',t')
\end{eqnarray}
\endnumparts
with initial conditions $\vec{r}\,'(t'=t)=\vec{r}$ and
$\vec{p}\,'(t'=t)=\vec{p}$. Note also that for the relativistic
Maxwellian, (\ref{eq:FDD}), we have
\begin{equation}
  \label{eq:df0sdp}
  \frac{\partial f_{0s}}{\partial \vec{p}} = - \frac{\Gamma_\mathrm{s} \, f_{0s}}
  {k_\mathrm{B}\,T_\mathrm{s}} \, \left( \vec{v} - U_s \, \vec{e}_\mathrm{y} \right)
\end{equation}
Expanding all physical scalar quantities~$\psi$ in terms of plane
waves with complex frequency~$\omega$ and real wavenumber
vector~$\vec{k}$ contained in the plane~(yOz)
\begin{equation}
  \label{eq:Dvlpt}
  \psi(\vec{r}, t) = \psi(x) \, \mathrm{exp} 
  (\rmi \, ( k_\mathrm{y} \, y + k_\mathrm{z} \, z - \omega\,t) )
\end{equation}
we arrive by standard techniques at expressions for the charge density
and current density
\numparts
\begin{eqnarray}
  \label{eq:Rho}
  \rho(\vec{r},t) & = & \sum_s 
  \frac{\Gamma_s \, \omega_{{\rm p}s}^2(\vec{r}) \, \varepsilon_0}{\Theta_s \, c^2} \, \left[
    U_s \, A_\mathrm{y}(\vec{r},t) - \phi(\vec{r},t) +
    \rmi \, ( \omega - k_\mathrm{y} \, U_s) \, \right. \times \nonumber \\
  & & \times \left.
    \int\!\!\!\int\!\!\!\int \hat{f}_{0s}(\vec{r}, \vec{p}) \, \int_{-\infty}^{\tau(t)}
    \left( \frac{\vec{p}\,'}{m_s} \cdot \vec{A}(\vec{r}\,',\tau') - 
      \gamma\,' \, \phi(\vec{r}\,',\tau') \right) \, \rmd\tau' \, \rmd^3\vec{p} \right] 
  \nonumber \\
 & & \\
  \label{eq:J}
  \vec{j}(\vec{r},t) & = & \sum_s 
  \frac{\Gamma_s \, \omega_{{\rm p}s}^2(\vec{r}) \, \varepsilon_0}{\Theta_s \, c^2} \, \left[
    ( U_s \, A_\mathrm{y}(\vec{r},t) - \phi(\vec{r},t) ) \, 
    \int\!\!\!\int\!\!\!\int \, \frac{\vec{p}\,c^2}{E} 
    \, \hat{f}_{0s}(\vec{r}, \vec{p}) \, \rmd^3\vec{p} 
    + \right. \nonumber \\
  & & + \rmi \, ( \omega - k_\mathrm{y} \, U_s) \,
    \int\!\!\!\int\!\!\!\int \, \frac{\vec{p}\,c^2}{E} \, 
    \hat{f}_{0s}(\vec{r}, \vec{p}) \, \times \nonumber \\
    & & \times \left. \int_{-\infty}^{\tau(t)}
    \left( \frac{\vec{p}\,'}{m_s} \cdot \vec{A}(\vec{r}\,',\tau') - 
      \gamma\,' \, \phi(\vec{r}\,',\tau') \right) \, \rmd\tau' \, \rmd^3\vec{p} \right]
\end{eqnarray}
\endnumparts
where the Lorentz factor of an unperturbed trajectory is $\gamma\,' =
\sqrt{ 1 + {p'}^2 / m_s^2 \, c^2}$.  Liouville's theorem
\begin{equation}
  \label{eq:Liouville}
  \hat{f}_{0s}(\vec{r}\,'(\tau'), \vec{p}\,'(\tau'), \tau') = 
  \hat{f}_{0s}(\vec{r}, \vec{p})  
\end{equation}
stating that $\hat{f}_{0s}$ is constant along the unperturbed
trajectories defined below by~(\ref{eq:intmvtr}) and
(\ref{eq:intmvtp}), was used to extract $\hat{f}_{0s}$ from the final
integration over~$\tau'$.  To compute the integrals, it is more
convenient to use the proper time defined by~$d\tau' = \rmd t' /
\gamma\,'$. This relation between proper and observer time also
defines the limit of time integration $\tau(t)$.  The
(non-relativistic) plasma frequency corresponding to species~''$s$''
is $\omega_{{\rm p}s}^2 = n_{0s} \, q_s^2 / m_s \, \varepsilon_0$ and
the normalised distribution function is defined by $f_{0s} = n_{0s} \,
\hat{f}_{0s}$.  The trajectories are integrated over the unperturbed
orbits, in the proper frame, following the equations of
motion~(\ref{eq:intmvtobsr}), (\ref{eq:intmvtobsp}): \numparts
\begin{eqnarray}
  \label{eq:intmvtr}
  \frac{\rmd\vec{r}\,'}{\rmd\tau'} & = & \frac{\vec{p}\,'(\tau')}{m_s} \\
  \label{eq:intmvtp}
  \frac{\rmd\vec{p}\,'}{\rmd\tau'} & = & \vec{p}\,'(\tau') \wedge \vec{\omega}_{Bs}(\vec{r}')
\end{eqnarray}
\endnumparts
with initial conditions $\vec{r} \,' (\tau'=\tau) = \vec{r}$ and
$\vec{p} \,' (\tau'=\tau) = \vec{p}$.  The non-relativistic cyclotron
frequency is given by $\vec{\omega}_{Bs} = q_s \, \vec{B}_0 / m_s$.
According to equation~(\ref{eq:intmvtr}) and~(\ref{eq:intmvtp}), both
energy $\gamma\,'$ and momentum component $p_\mathrm{z}'$ are
conserved so that the equations of motion in the z-direction can be
integrated analytically:
\numparts
\begin{eqnarray}
  \label{eq:intmvtz}
  z'(\tau') & = & z_0 + p_\mathrm{z} \, \tau' / m_s \\  
  \label{eq:intmvtpz}
  p_\mathrm{z}'(\tau') & = & p_\mathrm{z} = \mathrm{constant}
\end{eqnarray}
\endnumparts
Consequently, the $p_\mathrm{z}$ integration in~(\ref{eq:Rho})
and~(\ref{eq:J}) can be factored out and performed separately. After
some algebraic manipulations, the charge density is expressed as
\numparts
\begin{eqnarray}
  \label{eq:Rhox}
  \rho(x) & = & \sum_s \frac{n_{0s} \, q_s^2 \, \Gamma_s}{k_B \, T_s} \, \left[
    - \phi(x) + \rmi \, ( \omega - k_\mathrm{y} \, U_s) \, \right. \times \nonumber \\
  & & \left. \times \int\!\!\!\int_{R^2} 
    \hat{F}_{0s} (x,p_\mathrm{x},p_\mathrm{y}) \, S_\phi(x,p_\mathrm{x},p_\mathrm{y}) \,  
    \rmd p_\mathrm{x} \, \rmd p_\mathrm{y} \right]
\end{eqnarray}
and the current density as
\begin{eqnarray}
  \label{eq:Jxx}
  j_\mathrm{x}(x) & = & \sum_s \frac{n_{0s} \, q_s^2 \, \Gamma_s}{k_B \, T_s} \,
  \rmi \, ( \omega - k_\mathrm{y} \, U_s) \, \times \nonumber \\
  & & \times 
  \int\!\!\!\int_{R^2} \frac{p_\mathrm{x}}{m_s} \, \hat{F}_{0s} (x,p_\mathrm{x},p_\mathrm{y}) \, S_j(x,p_\mathrm{x},p_\mathrm{y}) \,  
  \rmd p_\mathrm{x} \, \rmd p_\mathrm{y} \\
  \label{eq:Jyx}
  j_\mathrm{y}(x) & = & \sum_s \frac{n_{0s} \, q_s^2 \, \Gamma_s}{k_B \, T_s} \, 
  \left[ \Gamma_s \, U_s^2 \, A_\mathrm{y}(x) +
    \rmi \, ( \omega - k_\mathrm{y} \, U_s) \, \right. \times \nonumber \\
  & & \left. \times 
    \int\!\!\!\int_{R^2} \frac{p_\mathrm{y}}{m_s} \, \hat{F}_{0s} (x,p_\mathrm{x},p_\mathrm{y}) \, S_j(x,p_\mathrm{x},p_\mathrm{y}) \,  
    \rmd p_\mathrm{x} \, \rmd p_\mathrm{y} \, \right] \\
  \label{eq:Jzx}
  j_\mathrm{z}(x) & = & \sum_s \frac{n_{0s} \, q_s^2 \, \Gamma_s}{k_B \, T_s} \,
  \rmi \, ( \omega - k_\mathrm{y} \, U_s) \, \times \nonumber \\
  &   &\times 
  \int\!\!\!\int_{R^2} \, \hat{F}_{0s} (x,p_\mathrm{x},p_\mathrm{y}) \, S_{j_\mathrm{z}}(x,p_\mathrm{x},p_\mathrm{y}) \,  
  \rmd p_\mathrm{x} \, \rmd p_\mathrm{y} 
\end{eqnarray}
\endnumparts
For brevity, we introduced the following functions~:
\numparts
\begin{eqnarray}
  \hat{F}_{0s} (x, p_\mathrm{x}, p_\mathrm{y}) & = & 
  \frac{\mathrm{exp} (\Gamma_s\,U_s\,p_\mathrm{y}/\Theta_s\,m_s\,c^2 ) }
  {4 \, \pi \, m_s^3 \, c^3 \, \Theta_s \, K_2(1/\Theta_s)} \\
  S_\phi(x, p_\mathrm{x}, p_\mathrm{y}) & = & \int_0^{-\infty} \, \left\{ \phi' \, I_{\rho_\phi} -
    \left[ ( p_\mathrm{x}' \, A_\mathrm{x}' + p_\mathrm{y}' \, A_\mathrm{y}' ) \, \frac{I_{\rho_1}}{m_s} + A_\mathrm{z}'
      \, I_{\rho_{A_\mathrm{z}}} \right] \right\} \, \rme^{\rmi \, \vec{k} \cdot \vec{r}'} \, d\tau' 
  \nonumber \\
& & \\
  S_j(x, p_\mathrm{x}, p_\mathrm{y}) & = & \int_0^{-\infty} \, \left\{ \phi' \, I_{\rho_1} -
    \left[ ( p_\mathrm{x}' \, A_\mathrm{x}' + p_\mathrm{y}' \, A_\mathrm{y}' ) \, \frac{I_{j_{A_1}}}{m_s} + A_\mathrm{z}'
      \, I_{j_{A_2}} \right] \right\} \, \rme^{\rmi \, \vec{k} \cdot \vec{r}'} \, d\tau' 
  \nonumber \\
& & \\
  S_{j_\mathrm{z}}(x, p_\mathrm{x}, p_\mathrm{y}) & = & \int_0^{-\infty} \, \left\{ \phi' \, I_{\rho_{A_\mathrm{z}}} -
    \left[ ( p_\mathrm{x}' \, A_\mathrm{x}' + p_\mathrm{y}' \, A_\mathrm{y}' ) \, \frac{I_{j_{A_2}}}{m_s} + A_\mathrm{z}'
      \, I_{j_{A_3}} \right] \right\} \, \rme^{\rmi \, \vec{k} \cdot \vec{r}'} \, d\tau'
  \nonumber \\
& & 
\end{eqnarray}
\endnumparts
We employ a short-hand notation for $(\phi', A_\mathrm{x}', A_\mathrm{y}', A_\mathrm{z}')$ which
should be understood as the scalar and vector potential evaluated at
$(\vec{r}\,', \tau')$ on the unperturbed
trajectories~(\ref{eq:intmvtr}) and~(\ref{eq:intmvtp}).  We also use
the following relation for the relativistic Maxwellian~(\ref{eq:FDD})
\begin{equation}
  \int\!\!\!\int\!\!\!\int \, \frac{\vec{p}\,c^2}{E} 
  \, \hat{f}_{0s} \, \rmd^3\vec{p} = \Gamma_s \, c \, \vec{\beta_s}
\end{equation}
It is easy to show that the proper time integration along the
unperturbed orbits in the z~direction, (\ref{eq:intmvtz}) and
(\ref{eq:intmvtpz}), leads to expressions involving modified Bessel
functions of the second kind $K_0$, $K_1$ and $K_2$ such that
\numparts
\begin{eqnarray}
  \label{eq:Integral}
  I_{\rho_1} & = & m_s \, c \, \gamma_\perp \, \left( \sqrt{\frac{\beta}{\alpha}} +
    \sqrt{\frac{\alpha}{\beta}} \right) \, K_1(2\,\sqrt{\alpha\,\beta}) \\
  I_{\rho_\phi} & = & m_s \, c \, \gamma_\perp^2 \, \left[ K_0(2\,\sqrt{\alpha\,\beta}) +
    \frac{1}{2} \, \left ( \frac{\beta}{\alpha} + \frac{\alpha}{\beta} \right) 
    \, K_2(2\,\sqrt{\alpha\,\beta}) \right] \\
  I_{\rho_{A_\mathrm{z}}} & = & m_s \, c^2 \, \gamma_\perp^2 \, 
  \frac{1}{2} \, \left ( \frac{\beta}{\alpha} - \frac{\alpha}{\beta} \right) 
  \, K_2(2\,\sqrt{\alpha\,\beta}) \\
  I_{j_{A_1}} & = & 2 \, m_s \, c \, K_0(2\,\sqrt{\alpha\,\beta}) \\
  I_{j_{A_2}} & = & m_s \, c^2 \, \gamma_\perp \, \left( \sqrt{\frac{\beta}{\alpha}} -
    \sqrt{\frac{\alpha}{\beta}} \right) \, K_1(2\,\sqrt{\alpha\,\beta}) \\
  \label{eq:Integralbis}
  I_{j_{A_3}} & = & m_s \, c^3 \, \gamma_\perp^2 \, \left[ 
    \frac{1}{2} \, \left ( \frac{\beta}{\alpha} + \frac{\alpha}{\beta} \right) 
    \, K_2(2\,\sqrt{\alpha\,\beta}) - K_0(2\,\sqrt{\alpha\,\beta}) \right]
\end{eqnarray}
\endnumparts
The argument of the modified Bessel functions for wave propagation
along~$\vec{e}_\mathrm{z}$ ($\vec{k} = k_\mathrm{z} \, \vec{e}_\mathrm{z}$) are given by~:
\begin{eqnarray}
  \label{eq:argument}
p_\perp & = & \sqrt{ p_\mathrm{x}^2 + p_\mathrm{y}^2 } \\
  \gamma_\perp & = & \sqrt{1 + p_\perp^2 / m_s^2 \, c^2} \\
  \alpha & = & \frac{\gamma_\perp}{2} \, \sqrt{ \Gamma_s/\Theta_s + 
    i \, ( \omega - k_\mathrm{z} \, c ) \, \tau' } \\
  \beta & = & \frac{\gamma_\perp}{2} \, \sqrt{ \Gamma_s/\Theta_s + 
    i \, ( \omega + k_\mathrm{z} \, c ) \, \tau' }
\end{eqnarray}

The integrals (\ref{eq:Integral})-(\ref{eq:Integralbis}) were first
derived by Trubnikov~\cite{1958PhDT........18T}. Note also that the
above integrals are computed assuming a relativistic Maxwellian
distribution function. The analytical integration over $p_{\rm z}$
leads to expressions involving modified Bessel functions $K_n$. In the
general case of non-Maxwellian distribution functions, an analytical
integration over $p_{\rm z}$ might not be possible, in which case it
would have to be performed numerically as for the $p_{\rm x,y}$
components. This will of course decrease the speed of computation of
the dispersion relation.

\section{The eigenvalue system}
\label{sec:eigenvalue}

\subsection{Derivation}

The eigenvalue system is found by solving the equations for the
electromagnetic potential determined according to the source
distribution given by~(\ref{eq:Rhox}), (\ref{eq:Jxx}),
(\ref{eq:Jyx}) and~(\ref{eq:Jzx}).  Inserting the latter expressions
into~(\ref{eq:OndePhi}) and~(\ref{eq:OndeA}), the eigenvalue system
reads~:
\numparts
\begin{eqnarray}
  \label{eq:SysPropPhiMono}
  \phi''(x) - \left( k^2 - \frac{\omega^2}{c^2} \right) \, \phi(x) +
  \frac{\rho(x)}{\varepsilon_0} & = & 0 \\
  \label{eq:SysPropAMono}
  \vec{A}''(x) - \left( k^2 - \frac{\omega^2}{c^2} \right) \, \vec{A}(x) + 
  \mu_0 \, \vec{j}(x) & = & 0
\end{eqnarray}
\endnumparts
where prime~$'$ denotes derivative with respect to~$x$.  The set of
equations~(\ref{eq:SysPropPhiMono}) and~(\ref{eq:SysPropAMono}) can be
written in terms of the unknown 4-dimensional vector $\vec{\Psi} =
(\phi, \vec{A})$
\begin{equation}
  \label{eq:SysProp}
  M(\omega,\vec{k}) \cdot \vec{\Psi} = 0
\end{equation}
It is a {\em non-linear} eigenvalue problem for the matrix~$M$ with
eigenvector $\vec{\Psi}$ and eigenvalue~$\omega$.
Daughton~\cite{1999PhPl....6.1329D} solved this system by locating the
zeroes of the determinant of~$M$.  We modify his method slightly by
solving simultaneously for both the eigenvalues and the eigenvectors.
A Newton-Raphson algorithm is used to find $\omega$ and $\vec{\Psi}$.
To check if the matrix~$M$ is singular, or in other words, if the
iteration process has converged, instead of computing the determinant
of~$M$, which should vanish, we attempt to zero the ratio between the
smallest and largest of the eigenvalues $\lambda_i$ of the matrix {\em
  linear} eigenvalue equation $M \cdot \vec{u} = \lambda \, \vec{u}$.
This latter eigenvalue equation should not be confused with our
original initial physical eigenvalue problem (\ref{eq:SysProp}) and is
absolutely not related to the computation of the dispersion relation.
This procedure of computing the eigenvalues~$\lambda$ of~$M$ is simply
another mean to verify if $M$ is singular. It helps to decide whether
the algorithm has converged or not. More details are given in the next
subsection.  This procedure allows us to track the dispersion curves
through crossing points. An example is given below.

\subsection{Algorithm}

Finding the roots of the determinant of the matrix~$M$ is the
traditional way to search for the {\em non-linear} eigenvalues in
(\ref{eq:SysProp}).  However, it is not the method of choice in our
problem because of the existence of different branches of the
dispersion relation. For multiple branches in the dispersion relation,
it is more efficient to look simultaneously for the eigenvalues and
the eigenvectors.

As an example, consider a homogeneous plasma (extension to an
inhomogeneous configuration is straightforward). In this case, the
matrix~$M$ is of dimension~$4\times4$, and $\vec{\Psi}$ has
4~components, one for $\phi$ and three for $\vec{A}$, each of which is
complex.
        
At this stage there are 10~unknowns:
\begin{itemize}
\item the complex eigenvalue~$\omega$ (real and imaginary parts)~;
\item the complex eigenfunction~$\vec{\Psi}$ (real and imaginary parts
  each of the 4~components).
\end{itemize}
We first reduce the system by normalising the eigenvector to unity,
$||\vec{\Psi}||=1$ and by setting its phase such that the imaginary
part of the last component of~ $\vec{\Psi}$ vanishes, which we refer
to as phase locking. The remaining 8~independent unknowns must then be
found by solving the nonlinear system~(\ref{eq:SysProp}), that consist
of 8~equations. To do this, we use the standard, globally convergent
Newton-Raphson method, \cite{1992nrca.book.....P}.

For fixed~$\vec{k}$, the iteration starts with an initial guess for
the eigenvalue $\omega = \omega_0$ and the eigenvector~$\vec{\Psi} =
\vec{\Psi}_0$.  An improved guess is found by Newton-Raphson iteration
in 8~dimensions.  Each step involves evaluating the Jacobian of the
system (\ref{eq:SysProp}) and uses line searches and backtracking as
described in \cite{1992nrca.book.....P}. The iteration is stopped when
the matrix~$M$ is close to being singular, i.e.  when $\mathrm{det} \,
M \approx 0$.  This condition is verified indirectly by computing the
four complex eigenvalues of~$M$, $(\lambda_1, \lambda_2, \lambda_3,
\lambda_4)$, that are solutions of the {\em linear} eigenvalue problem
$M \cdot \vec{u} = \lambda \, \vec{u}$.  We emphasize that
the~$\lambda_i$ should not be confused with the eigenvalues~$\omega$
of our original {\em nonlinear} problem.  When the ratio of the
smallest to the largest~$\lambda_i$ is sufficiently small, namely when
\begin{equation}
  \label{eq:CV}
  \frac{\mathrm{min}_{i\,\in\,[1..4]} \, || \lambda_i ||}
  {\mathrm{max}_{i\,\in\,[1..4]} \, || \lambda_i ||} < \varepsilon 
\end{equation}
with, typically, $\varepsilon = 10^{-6}...10^{-8}$, we assume that
convergence has been achieved and terminate the iteration procedure.

\section{Results}
\label{sec:results}

We apply our algorithm to the following cases~:
\begin{itemize}
\item the familiar electromagnetic oscillation modes of a
  nonrelativistic magnetised plasma~;
\item a counter-streaming pair-plasma with nonrelativistic temperature
  and nonrelativistic drift speed corresponding to two oppositely
  shifted Maxwellian distribution functions~;
\item a counter-streaming pair plasma with relativistic temperature
  and drift speeds close to the velocity of light, $U_s=0.05-0.3~c$.
\end{itemize}

\subsection{Plasma oscillations}

\begin{figure}[htbp]
  \centering
  \includegraphics[width=10cm]{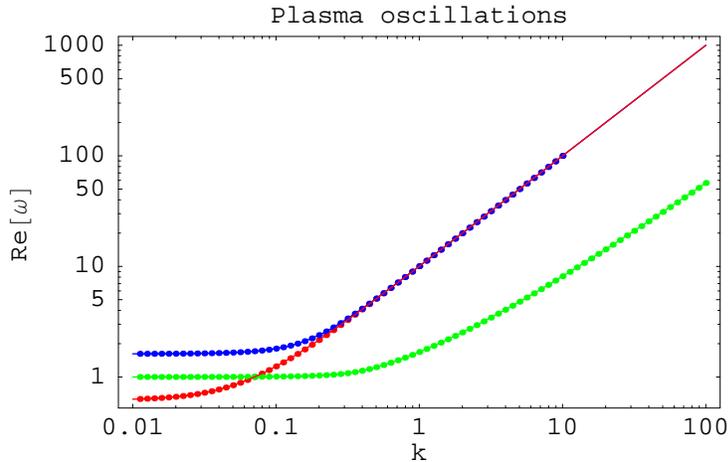}
  \caption{Dispersion relation for the transverse and 
    longitudinal modes in a non-relativistic magnetised plasma. The
    exact analytical expressions for electrostatic waves are shown in
    green curve and those for electromagnetic waves are shown in red
    and blue curves. The plasma frequency is normalised to unity.}
  \label{fig:ES}
\end{figure}

The dispersion relation for a single species plasma in a homogeneous
magnetic field obtained using a nonrelativistic version of our code is
shown in figure~\ref{fig:ES}. The plasma frequency is normalised to
unity~$\omega_{\rm p}=1$ and the thermal speed of the non-relativistic
Maxwellian is~$v_{\rm th} = 0.1\,c$. In figure~\ref{fig:ES}, we compare
our numerical results with the analytic expressions found using a
simple root finding algorithm from the exact dispersion relation for
wave propagation parallel to the magnetic field~$\vec{B}_0$, ($k_\mathrm{y}=0$)
in order to check the correctness of our algorithm implementation.
For electrostatic waves, we found the usual dispersion relation
\begin{equation}
  \label{eq:OndesES}
  1 + 2 \, \frac{\omega_\mathrm{p}^2}{k_\mathrm{z}^2 \, v_\mathrm{th}^2} \, 
  \left[ 1 + \frac{\omega}{k_\mathrm{z} \, v_\mathrm{th}} \, 
    Z\left( \frac{\omega}{k_\mathrm{z} \, v_\mathrm{th}} \right) \right] = 0
\end{equation}
where the plasma dispersion function~$Z$ is defined for
$\mathrm{Im}(\zeta)>0$ by
\begin{equation}
  \label{eq:FonctionZ}
  Z(\zeta) = \frac{1}{\sqrt{\pi}} \, \int_{-\infty}^{+\infty}
  \frac{\rme^{-t^2}}{t - \zeta} \, \rmd t
\end{equation}
and analytically continued for $\mathrm{Im}(\zeta)<0$, see for
instance Delcroix and Bers~\cite{1994Delcroix}.  For the left and
right-handed circularly polarised electromagnetic wave we found
\begin{equation}
  \label{eq:OndesEM}
  \frac{k_\mathrm{z}^2 \, c^2}{\omega^2} =
  1 + \frac{\omega_\mathrm{p}^2}{\omega \, k_\mathrm{z} \, v_\mathrm{th}} \, 
  Z\left( \frac{\omega \pm \omega_B}{k_\mathrm{z} \, v_\mathrm{th}} \right)
\end{equation}

The numerical results are compared with the analytical exact
expression for electrostatic waves (green curve, (\ref{eq:OndesES}))
and electromagnetic waves (red and blue curves, (\ref{eq:OndesEM})).
The advantages of implementing our modified root finding algorithm by
looking for eigenvalues simultaneously with eigenfunctions is evident.
The algorithm has no difficulty tracking only one branch of the
dispersion relation even when crossing another branch. However it
requires more computational effort since the root finding occurs in a
space of higher dimension than is needed to find the eigenvalue alone.

\begin{figure}[htbp]
  \centering
  \includegraphics[width=10cm]{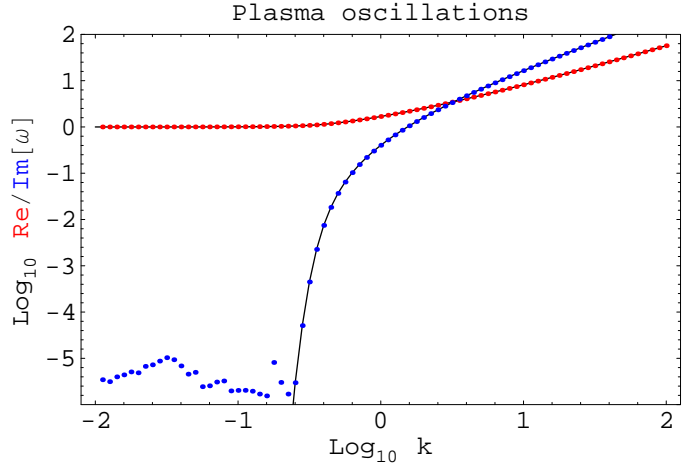}
  \caption{Dispersion relation for electrostatic plasma 
    waves. The real and imaginary parts of the eigenfrequency are
    shown in red and blue dots respectively. The plasma frequency is
    normalised to unity.}
  \label{fig:ESIm}
\end{figure}

To test our algorithm in cases where the imaginary part~${\rm
  Im}(\omega)$ is small compared to the real part~${\rm Re}(\omega)$,
we consider electrostatic plasma oscillations, in the short wavelength
limit, $k\ll1$.  Numerical results for ${\rm (Re/Im)}(\omega)$,
respectively red and blue dots, are compared with the analytical
dispersion relation, solid black curves, (\ref{eq:OndesES}),
figure~\ref{fig:ESIm}. As long as the ratio ${\rm Im}(\omega)/{\rm
  Re}(\omega)$ is larger than the accuracy, which is about 5~digits,
the results are reliable.

\subsection{``Classical'' Weibel instability}

Next we check our algorithm for a situation with two counter-streaming
species of non-relativistic temperature, taking $v_\mathrm{th}/c =
10^{-3}$ and a small drift speed, $U_s = 0.1\,c$.

\begin{figure}[htbp]
  \centering
  \includegraphics[width=10cm]{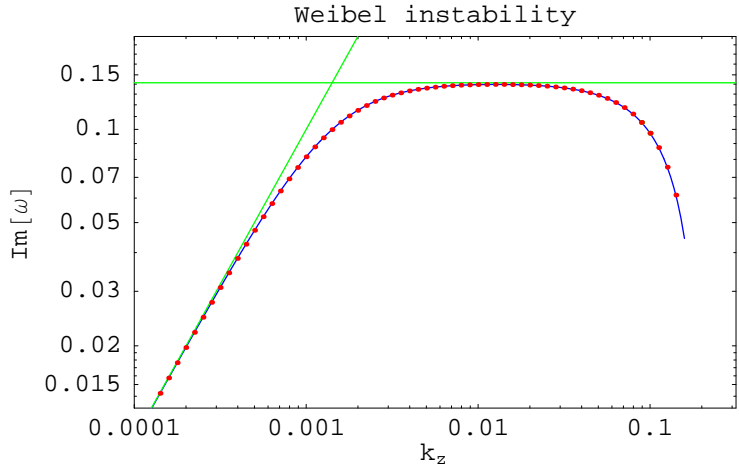}
  \caption{Growth rate for the non-relativistic Weibel 
    instability~($\Theta_s\ll1$). Results from our algorithm are
    depicted with red points, the exact analytical expression in blue
    curve and the asymptotic matching in case of a zero temperature
    plasma~($\Theta_s=0$) in green lines.}
  \label{fig:Weibel1}
\end{figure}

We recall that the exact analytical dispersion relation for the
classical counter-streaming Weibel instability for equal and opposite
drift speed for both species with the same plasma frequency
$\omega_{\rm p}$ is given by
\begin{equation}
  \label{eq:WeibelExact}
  1 - \frac{k_\mathrm{z}^2 \, c^2}{\omega^2} - 2 \, \frac{\omega_{\rm p}^2}{\omega^2} \,
  \left[ 1 - \left( 1 + 2 \, \frac{U_s^2}{v_\mathrm{th}^2} \right) \, 
    \left( 1 + \frac{\omega}{k_\mathrm{z} \, v_\mathrm{th}} \, 
      Z\left( \frac{\omega}{k_\mathrm{z} \, v_\mathrm{th}} \right) \right) \right] = 0 
\end{equation}
The results are adapted from Delcroix and Bers~\cite{1994Delcroix}.
In figure~\ref{fig:Weibel1} we show the dispersion relation for the
Weibel instability (red points) compared with the analytical
expression (blue curve, (\ref{eq:WeibelExact})) and the asymptotic
matching in case of a zero temperature plasma (green lines). The
agreement is excellent.

\begin{figure}[htbp]
  \centering
  \includegraphics[width=14cm]{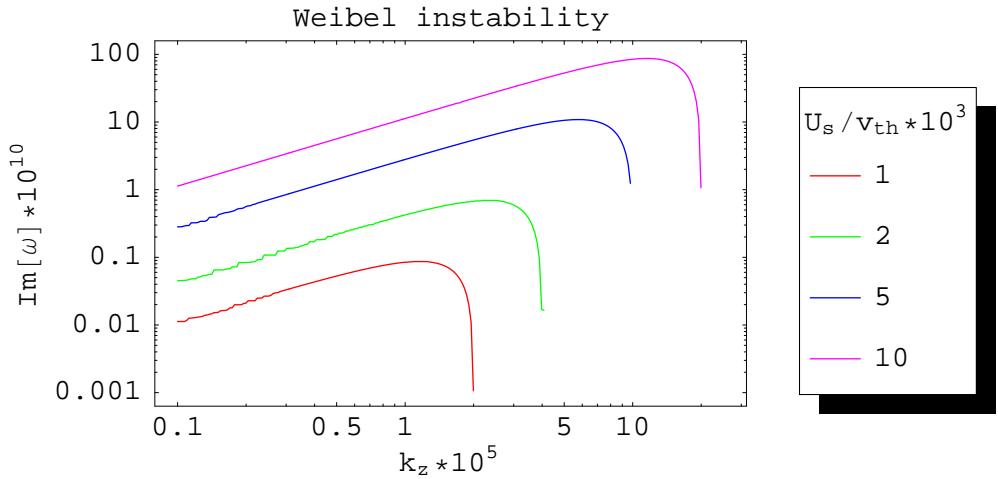}
  \caption{Growth rate for the non-relativistic Weibel 
    instability~($\Theta_s\ll1$) for different drift speeds~$U_s$.}
  \label{fig:Weibel2}
\end{figure}

Results for the dispersion relation in the case of small drift speeds
are also shown in figure~\ref{fig:Weibel2}.  Here, the large
$k_\mathrm{z}$ asymptote is absent due to thermal effects.

\subsection{Relativistic Weibel instability}

\begin{figure}[htbp]
  \centering
   \includegraphics[width=14cm]{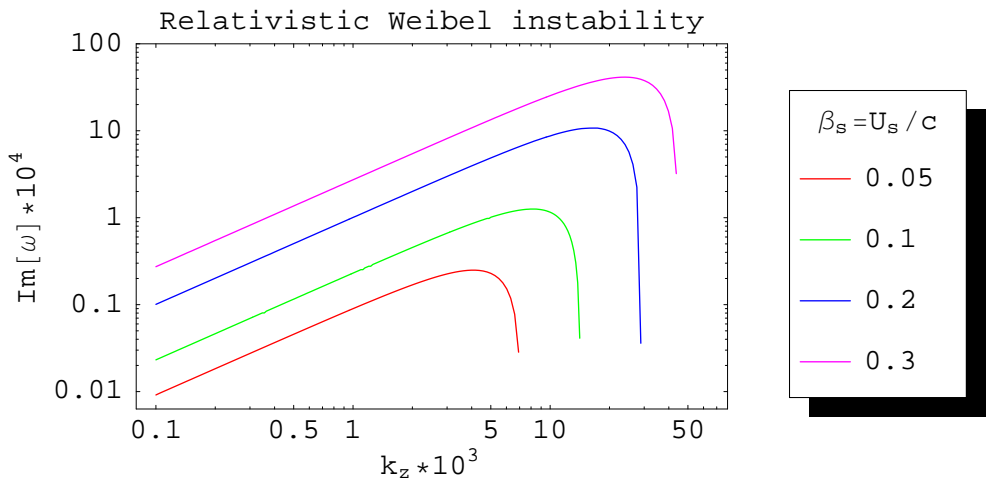}
   \caption{Growth rate for the relativistic Weibel instability
     ($\Theta_s\gg1$) for different relativistic drift speeds~$U_s$.}
  \label{fig:Weibel3}
\end{figure}

Finally we computed the dispersion relation for a two-component fully
relativistic counter-streaming plasma with high temperature, $\Theta_s
= 10^2$ and different drift speeds as shown in
figure~\ref{fig:Weibel3}.  The dispersion curves are similar to those
obtained in the classical Weibel instability.  Due to the spread in
momentum present at finite temperature, the instability is suppressed
for large wavenumbers~$k_\mathrm{z}$ in both the classical and
relativistic cases.

\section{Conclusion}
\label{sec:conclusion}

We have constructed an algorithm to solve the dispersion relation for
non-relativistic and relativistic multi-component plasmas. The code
has been validated by comparing the results with some typical
configurations with homogeneous magnetic field, for which the
analytical dispersion relation is known. New results have been
obtained for the growth rate of the relativistic Weibel or two-stream
instability for two counter-streaming relativistic Maxwellians which
complement those found using a water-bag distribution.  This code is
easily extended to inhomogeneous plasmas, and is therefore a suitable
tool for the study of stability properties of configurations of
interest in gamma-ray burst and pulsar wind theories.

\ack{This research was supported by a grant from
the G.I.F., the German-Israeli Foundation for Scientific Research and
Development.}

%\bibliography{/home/petri/pulsar/rapport/bibliotot}

\section*{References}

\begin{thebibliography}{10}

\bibitem{2005RvMP...76.1143P}
T.~{Piran}.
\newblock {The physics of gamma-ray bursts.}
\newblock {\em Reviews of Modern Physics}, 76:1143--1210, 2005.

\bibitem{2005RMxAC..23...27M}
F.~C. {Michel}.
\newblock {Winds from Pulsars}.
\newblock In {\em Revista Mexicana de Astronomia y Astrofisica Conference
  Series}, pages 27--34, October 2005.

\bibitem{2005PPCF...47B.719K}
J.~G. {Kirk}.
\newblock {Relativistic plasmas in pulsar winds}.
\newblock {\em Plasma Physics and Controlled Fusion}, 47:B719--B726, December
  2005.

\bibitem{1999PhPl....6.1329D}
W.~{Daughton}.
\newblock {The unstable eigenmodes of a neutral sheet}.
\newblock {\em Physics of Plasmas}, 6:1329--1343, April 1999.

\bibitem{1959PhRvL...2...83W}
E.~S. {Weibel}.
\newblock {Spontaneously Growing Transverse Waves in a Plasma Due to an
  Anisotropic Velocity Distribution}.
\newblock {\em Physical Review Letters}, 2:83--84, February 1959.

\bibitem{1987PhRvA..35.2718Y}
P.~H. {Yoon} and R.~C. {Davidson}.
\newblock {Exact analytical model of the classical Weibel instability in a
  relativistic anisotropic plasma}.
\newblock {\em \pra}, 35:2718--2721, March 1987.

\bibitem{1989PhFlB...1.1336Y}
P.~H. {Yoon}.
\newblock {Electromagnetic Weibel instability in a fully relativistic
  bi-Maxwellian plasma}.
\newblock {\em Physics of Fluids B}, 1:1336--1338, June 1989.

\bibitem{2002PhPl....9.2458S}
L.~O. {Silva}, R.~A. {Fonseca}, J.~W. {Tonge}, W.~B. {Mori}, and J.~M.
  {Dawson}.
\newblock {On the role of the purely transverse Weibel instability in fast
  ignitor scenarios}.
\newblock {\em Physics of Plasmas}, 9:2458--+, June 2002.

\bibitem{2004A&A...428..365W}
J.~{Wiersma} and A.~{Achterberg}.
\newblock {Magnetic field generation in relativistic shocks. An early end of
  the exponential Weibel instability in electron-proton plasmas}.
\newblock {\em \aap}, 428:365--371, December 2004.

\bibitem{2006ApJ...647.1250L}
Y.~{Lyubarsky} and D.~{Eichler}.
\newblock {Are Gamma-Ray Burst Shocks Mediated by the Weibel Instability?}
\newblock {\em \apj}, 647:1250--1254, August 2006.

\bibitem{1993PhFlB...5.3369Y}
T.-Y.~B. {Yang}, Y.~{Gallant}, J.~{Arons}, and A.~B. {Langdon}.
\newblock {Weibel instability in relativistically hot magnetized
  electron-positron plasmas}.
\newblock {\em Physics of Fluids B}, 5:3369--3387, September 1993.

\bibitem{1982AuJPh..35...41M}
D.~B. {Melrose}.
\newblock {Covariant description of dispersion in a relativistic thermal
  electron gas}.
\newblock {\em Australian Journal of Physics}, 35:41--+, 1982.

\bibitem{2004PhPl...11g5532S}
R.~{Schlickeiser}.
\newblock {Covariant kinetic dispersion theory of linear waves in anisotropic
  plasmas. I. General dispersion relation, bi-Maxwellian distribution and non
  relativistic limits}.
\newblock {\em Physics of Plasmas}, 11:5532--+, December 2004.

\bibitem{2006PhPl...13b2107S}
U.~{Schaefer-Rolffs} and R.~{Schlickeiser}.
\newblock {The relativistic kinetic Weibel instability: General arguments and
  specific illustrations}.
\newblock {\em Physics of Plasmas}, 13:2107, 2006.

\bibitem{2005PhPl...12.2901T}
R.~C. {Tautz} and R.~{Schlickeiser}.
\newblock {Counterstreaming magnetized plasmas. I. Parallel wave propagation}.
\newblock {\em Physics of Plasmas}, 12:2901--+, December 2005.

\bibitem{2006PhPl...13f2901T}
R.~C. {Tautz} and R.~{Schlickeiser}.
\newblock {Counterstreaming magnetized plasmas. II. Perpendicular wave
  propagation}.
\newblock {\em Physics of Plasmas}, 13:2901--+, June 2006.

\bibitem{2002PhPl....9.1104S}
I.~{Silin}, J.~{B{\"u}chner}, and L.~{Zelenyi}.
\newblock {Instabilities of collisionless current sheets: Theory and
  simulations}.
\newblock {\em Physics of Plasmas}, 9:1104--+, April 2002.

\bibitem{1970PhRvA...1.1772B}
B.~{Buti}.
\newblock {Superluminous Waves in Streaming Relativistic Plasmas}.
\newblock {\em \pra}, 1:1772--1774, June 1970.

\bibitem{1972PhRvA...5.1558B}
B.~{Buti}.
\newblock {Streaming Instabilities in Relativistic Magnetoplasmas}.
\newblock {\em \pra}, 5:1558--1563, March 1972.

\bibitem{1979SvA....23..460Z}
L.~M. {Zelenyi} and V.~V. {Krasnoselskikh}.
\newblock {Relativistic Modes of Tearing Instability in a Background Plasma}.
\newblock {\em Soviet Astronomy}, 23:460--+, August 1979.

\bibitem{1958PhDT........18T}
B.~A. {Trubnikov}.
\newblock {\em {Magnetic Emission of High Temperature Plasma}}.
\newblock PhD thesis, Dissertation, Moscow (US-AEC Tech.~Inf.~Service,
  AEC-tr-4073 [1960]), (1958), 1958.

\bibitem{1992nrca.book.....P}
W.~H. {Press}, S.~A. {Teukolsky}, W.~T. {Vetterling}, and B.~P. {Flannery}.
\newblock {\em {Numerical recipes in C. The art of scientific computing}}.
\newblock Cambridge: University Press, |c1992, 2nd ed., 1992.

\bibitem{1994Delcroix}
J.L. {Delcroix} and A.~{Bers}.
\newblock {\em {Physique des plasmas - Tome 2}}.
\newblock EDP Sciences - CNRS Editions, 1994.

\end{thebibliography}
%\bibliographystyle{unsrt}

\end{document}